# Interplay between Topological States and Rashba States as Manifested on Surface Steps at Room Temperature

Wonhee Ko[1,3,†], Seoung-Hun Kang[2,†], Jason Lapano[2], Hao Chang[1,3], Jacob Teeter[1], Hoyeon Jeon[1], Matthew Brahlek[2], Mina Yoon[2,*], Robert G. Moore[2,*], An-Ping Li[1,3,*]

[1]Center for Nanophase Materials Sciences, Oak Ridge National Laboratory, Oak Ridge, Tennessee 37831, USA

[2]Materials Science and Technology Division, Oak Ridge National Laboratory, Oak Ridge, Tennessee 37831, USA

[3]Department of Physics and Astronomy, The University of Tennessee, Knoxville, Tennessee 37996, USA

[†]These authors contributed equally to this work

Email: myoon@ornl.gov, moorerg@ornl.gov, apli@ornl.gov

**Abstract**

The unique spin texture of quantum states in topological materials underpins many proposed spintronic applications. However, realizations of such great potential are stymied by perturbations, such as temperature and local fields imposed by impurities and defects, that can render a promising quantum state uncontrollable. Here, we report room-temperature observation of interaction between Rashba states and topological surface states, which manifests unique spin textures controllable by layer thickness of thin films. Specifically, we combine scanning tunneling microscopy/spectroscopy with the first-principles theoretical calculation to find the robust Rashba states coexisting with topological surface states along the surface steps with characteristic spin textures in momentum space. The Rashba edge states can be switched off by reducing the thickness of a topological insulator $Bi_2Se_3$ to bolster their interaction with the hybridized topological surface states. The study unveils a manipulating mechanism of the spin textures at room temperature, reinforcing the necessity of thin film technology in controlling quantum states.

**Keywords**

Topological insulators, $Bi_2Se_3$ film, Topological surface states, Rashba edge states, Scanning tunneling microscopy, First-principles density functional theory, Tight-binding modeling

In topological materials with large spin-orbit coupling, several quantum states possess unique spin texture to be useful for spintronic functionality. For example, in three-dimensional (3D) quantum spin Hall insulators (i.e. topological insulators), topological surface states exhibit spin-momentum locking [1,2] which allows conversion of charge current to spin polarization [3,4] or vice versa [5,6]. Topological insulators also host Rashba states on the surface caused by inversion symmetry breaking and quantum confinement [7,8], although topologically trivial, whose spin texture induces spin-charge conversion through Rashba-Edelstein effect [9,10]. The joint effect of the two spin-related interactions can lead to complex spin and charge transport phenomena that remain to be evaluated [11].

The spintronic functionality, however, of the topological insulator-based devices has been mostly demonstrated at the cryogenic temperature [4,12-16]. One reason is that even though the quantum states are protected by non-trivial topology and manifested themselves at room temperature [7,17-19], their electronic properties are significantly hampered by external perturbations, such as unintentional doping, thermally excited bulk carriers, and intrinsic defects [20-25]. A representative example of such deterioration is quantum transport in $MnBi_2Te_4$, where the combined magnetism and non-trivial topology are expected to result in ballistic spin transport by quantum anomalous Hall effect at relatively high temperature [26-28], but its experimental realization is still controversial [29-32] . Moreover, manipulating these quantum states involves the interaction in the energy scale of typically order of magnitude less than the state itself [9,33], further restricting the demonstration of any device functionality at ambient condition. To achieve practical spintronic devices with topological materials, it is critical to realize robust quantum states with spin texture and well-defined interaction.

In this study, we show that the surface steps of topological insulators create robust Rashba edge states and demonstrate their interactions with the topological surface states, all at room temperature. Specifically, we investigated $Bi_2Se_3$ thin film with precisely tuned nominal thickness and used scanning tunneling microscopy/spectroscopy (STM/S) to identify the single quintuple layer (QL) steps and characterize their electronic structure [Fig. 1(a)]. For the films thicker than critical QL thickness $N_c$, steps induce translational symmetry breaking, giving rise to the Rashba edge states that are detected as the enhanced differential tunneling conductance localized along the steps at around Dirac point $E_D$. The Rashba edge states are robust against edge geometry and dislocations, and persist up to the room temperature. However, the edge states disappear in STM/S when the film thickness becomes less than $N_c$ QL, where $N_c = 7$~9. The first-principles density functional theory (DFT) calculations confirm the coexistence of topological surface states and Rashba edge states at the surface steps with distinct band dispersion and spin texture for the films thicker than 5 QL. Furthermore, theoretical calculations show that the top and bottom topological surface states start to hybridize and overlap significantly with the Rashba edge states when the film thickness is reduced below 5 QL, whose value is close to the $N_c$ from STM observation, and as a result, the edge states become delocalized across the surface. The observation is an unambiguous demonstration of the interaction between the topological and Rashba states at room temperature.

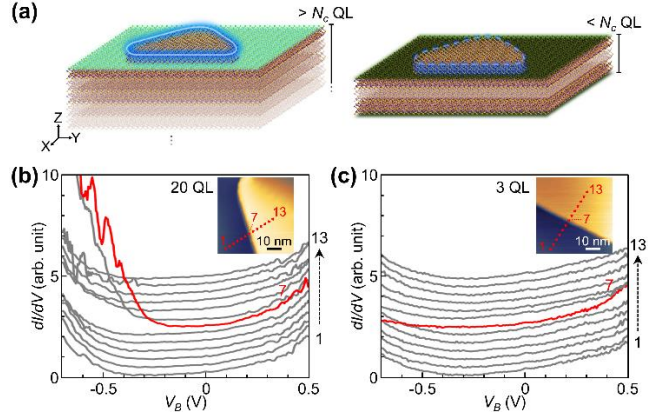

**FIG. 1. Rashba edge states in $Bi_2Se_3$ film and its interplay with topological surface states.** (a) Schematic of the interplay between Rashba edge states and topological surface states for the films thicker than critical QL thickness $N_c$ (right) and thinner than $N_c$ (left). In the thick film (> $N_c$ QL), bright green surface and blue line indicates topological surface states and Rashba edge states, respectively. However, in the thin film (< $N_c$ QL), the topological surface states open the gap (dark green surface), and edge states disappear (dotted blue line). (b,c) $dI/dV$ spectra across the step taken on 20 QL film and 3 QL film, respectively. Inset of (b) and (c) shows the topographic image of the film ($V_B$ = 1 V, $I$ = 1 pA for (b); $V_B$ = 1 V, $I$ = 2 pA for (c)). The locations of spectra are marked by red dots and numbers on the topographic images.

To characterize the edge states and observe their interaction with topological surface states, we took the $dI/dV$ spectra across the single QL steps in the $Bi_2Se_3$ thin films with varying thicknesses. Specifically, 20 QL and 3 QL films were investigated to compare the behavior of edge states for thickness above and below the critical thickness of 5 QL, which was reported as the thickness that starts to open the band gap at the Dirac point by hybridization of the top and bottom surface states [34]. The ARPES measurement confirmed that 20 QL film retains the crossing of bands at the Dirac point of topological surface states, while 3 QL film displays the clear band gap opening at the Dirac point [Fig. S1] [35]. The large- and atomic-scale STM images of both films display atomically flat surface with a few single QL step edges [Fig. S2]. In Fig. 1(b) and (c), we compare *dI/dV* spectra across the single QL step edges on the films with a nominal thickness of 20 QL and 3 QL, respectively. At the terraces far from the edges, the *dI/dV* spectra of both films display a similar dip shape with the minimum at about -0.3 eV, which is consistent with the existence of the surface states with Dirac point $E_D \sim -0.3$ eV in $Bi_2Se_3$ [36,37]. However, when the tip is placed right on top of the step, *dI/dV* spectra (red curve) from the 20 QL sample exhibit a significant enhancement of intensity below $E_D$. Such an increase indicates the existence of the edge states localized along the physical step for energy from $E_D$ to the bulk valence band, which was previously predicted and observed on topological insulators [38-40]. In striking contrast, the *dI/dV* spectra from the 3 QL sample do not show any enhancement in the energy range of -0.7 ~ 0.3 eV across the step edge, and the spectrum right above the step edge even displays a slight decrease in the intensity below $E_D$. The results strongly indicate that the signature of the edge states disappears when the film thickness is reduced to 3 QL. Further measurement of films with various thicknesses, different substrates, and temperatures of 4.6 K and room temperatures, consistently exhibited the disappearance of edge states below 6 QL and appearance of edge states above 10 QL [Fig. S3].

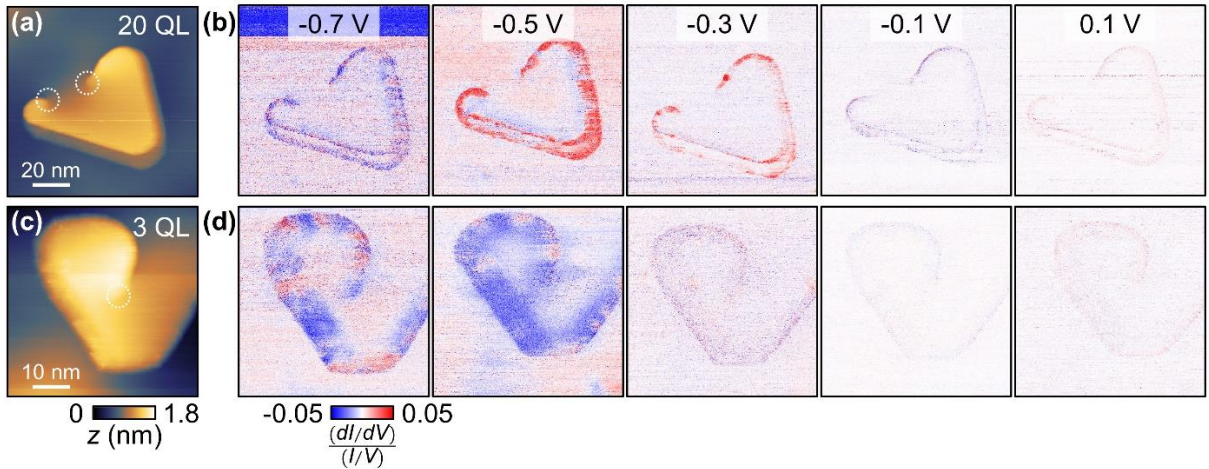

**FIG. 2. Normalized d*I*/d*V* maps of a single QL steps on the 20 QL and 3QL samples.** (a) A topographic image of single QL step on 20 QL film. ($V_B = 0.3$ V, $I = 100$ pA). Dotted circles denote the screw dislocations. (b) Normalized *dI/dV* maps taken at the same area of (a) for different bias voltages. (c) A topographic image of single QL step on 3 QL film ($V_B$ = -0.7 V, $I$ = 100 pA). Dotted circles denote the screw dislocations. (d) Normalized $dI/dV$ maps taken at the same area of (c) for different bias voltages.

To investigate the spatial distribution of the edge states and their dependence on the film thickness, we measured $dI/dV$ maps at the region with single QL steps for 20 QL and 3 QL films [Fig. 2]. To remove the artifacts in $dI/dV$ maps from the fluctuation of current, we normalized $dI/dV$ by dividing it with $I/V$ [Fig. S4] [35]. When $E > E_D$, there is little contrast along the edges for both 20 QL and 3 QL films, indicating that no edge states exist in this energy range. However, for 20 QL film, $dI/dV$ maps at the $E_D$ and below (-0.5 ~ -0.3 eV) show significant enhancement of intensity along the step edge [Fig. 2(b)]. This confirms the existence of edge states that are tightly localized at the physical step, which is consistent with the observation in Fig. 1(b) that displays the new states below $E_D$ for $dI/dV$ spectra at the edge. Moreover, the $dI/dV$ maps show that the edge states are robust against the geometrical disorders, as they persist over the curves and screw dislocations in the step edge. The $dI/dV$ maps on 3 QL film, however, display drastically different features. For all energies, including $E_D$, $dI/dV$ along the step edges do not show discernable enhancement [Fig. 2(d)]. At $E_D$ and below (-0.5 ~ -0.3 eV), $dI/dV$ signal displayed slight depression in contrast. No appearance of enhanced $dI/dV$ along the step edge verifies that the edge states have disappeared for the 3 QL film. When the energy reaches deep inside the bulk valence band (-0.7 eV), both 20 QL and 3 QL films exhibit a similar pattern of depression along the edges. The observation indicates that the disappearance of the edge state signal is most likely related to the topological surface states.

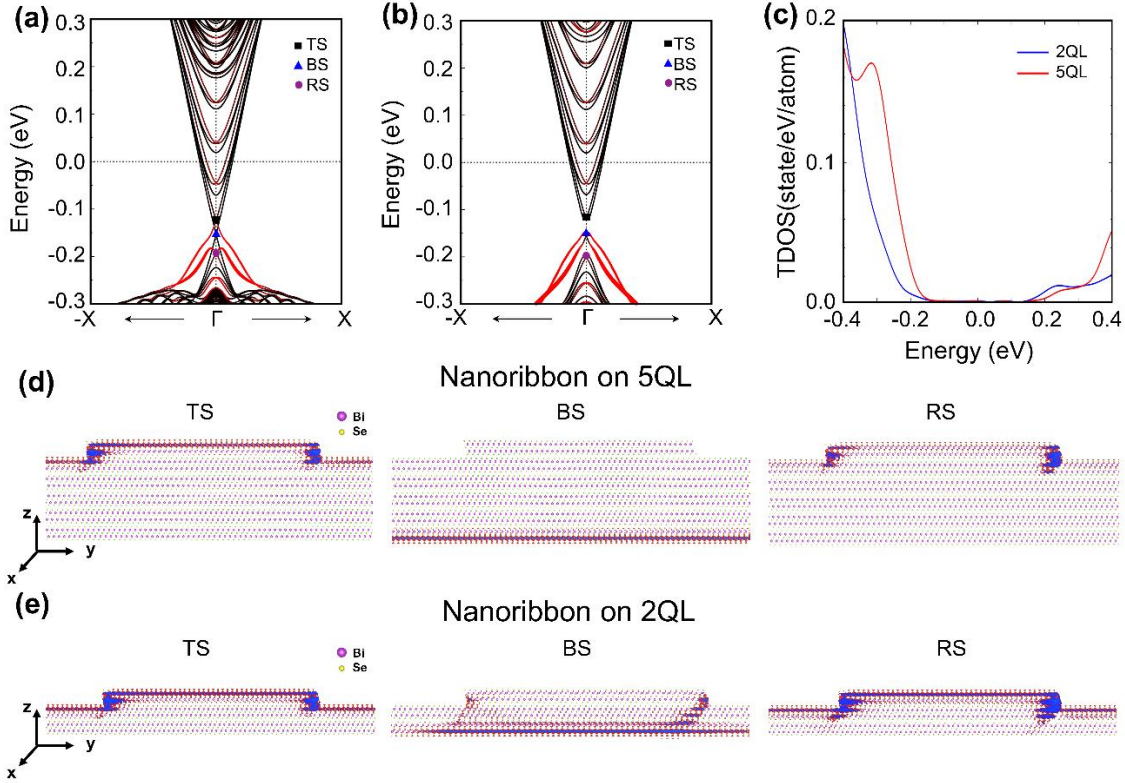

**FIG. 3. Theoretical electronic structures of nanoribbon on thick (5QL) and thin (2QL) layers.** (a) Band structure for nanoribbon on thick and (b), thin layers. The red color is proportional to the band projection at the step edge. (c) Total density of states of nanoribbon on thick (red line) and thin (blue line) layers. (d) Charge density at the gamma for top, bottom surface states (TS, BS), and Rashba states (RS) denoted by a black square, blue triangle, and purple circle dots in (a), respectively. (e) Charge density at the gamma for TS, BS, and RS denoted by a black square, blue triangle, and purple circle dots in (b), respectively.

To understand the origins of localized edge states and their thickness dependence, we investigated the topological surface states of $Bi_2Se_3$ according to the film thickness using the reliable tight-binding parameters of $Bi_2Se_3$. To confirm the strength of crosstalk between the top and bottom topological surface states, we checked the band gap as the function of thickness. As the film thickness increases from 1 QL, the band gap gradually decreases until the band crossing at the Dirac point is confirmed in 5 QL [Fig. S5]

[34,35]. Experimentally, the strength of crosstalk between the top and bottom topological surfaces can be estimated by the band gap measured by ARPES. The band gap of 3 QL film is about 100 meV [Fig. S1(b)] [35], which is in good agreement with 2 QL in our simulation [Figs. S5(c) and (d)] [35]. Therefore, we compare the thick (20 QL) and thin (3 QL) film case of the experiment to the electronic structure simulation of the nanoribbons on 5 QL and 2QL films. Figure 3(a) shows the additional states localized at the step edge in the 5 QL film. The pristine topological insulator without any step edge should present the Dirac point with a degenerated linear dispersion from both the top and bottom surface states around $E_D$. However, since the step breaks the inversion symmetry, the bands of top and bottom Dirac states become split, and 1D localized edge states are formed, as shown in Figs. 3(a) and (d). The 1D localized edge states are the type-1 Rashba state, i.e., one possesses a momentum-dependent splitting of the spin band induced by strong spin-orbit coupling (SOC) and asymmetric potential [Fig. S7] [35]. Further calculation of 10 QL film displays the same 1D Rashba states although the bulk states change [Fig. S10]. The energy and spatial distribution of the 1D Rashba states in Figs. 3(a) and (d) are consistent with the edge states observed in the *dI/dV* maps of 20 QL film in Fig. 2(b).

We further confirmed the robustness of Rashba states against variations in step edge shapes by conducting a comprehensive tight-binding study on various edge configurations [Fig. S8, S9 and Table S2]. Our results revealed that, under the Se-rich condition of our MBE synthesis, all edge terminations including ziazag, armchair, and triangular types harbor the Rashba edge states with a substantial Rashba strength. We also note that although some topological materials display topological edge states in 2D limit [41,42], the observed edge states cannot be topological since our theoretical topological index ($Z_2$) calculation shows that $Bi_2Se_3$ films have trivial topology ($Z_2 = 0$) up to 8 QL.

In contrast, the thin film (2 QL) does not form localized 1D Rashba state and band crossing around $E_D$ [Fig. 3(b)]. Instead, both top surface states and Rashba states from the nanoribbon edges interact with bottom surface state, resulting in coupled and gapped states as shown in Figs. 3(b) and (e). So, the coupling between Rashba states and bottom surface states render delocalized hybridized states, as shown in Fig. 3(e), which explains the disappearance of *dI/dV* features from the edge states for 3QL film in Fig. 2(d). We note that the interaction between Rashba state and top surface states does not appear near the gamma point due to the different energy levels as shown in Figs. 3(a) and (b). The calculation also confirms the difference in the total density of states below the -0.2 eV in Fig. 3(c), which agrees with the thickness-dependent appearance of a peak below $E_D$ in $dI/dV$ spectra at the edge [Figs. 1(b) and (c)]. Therefore, our theoretical simulations agree well with the experimental observations.

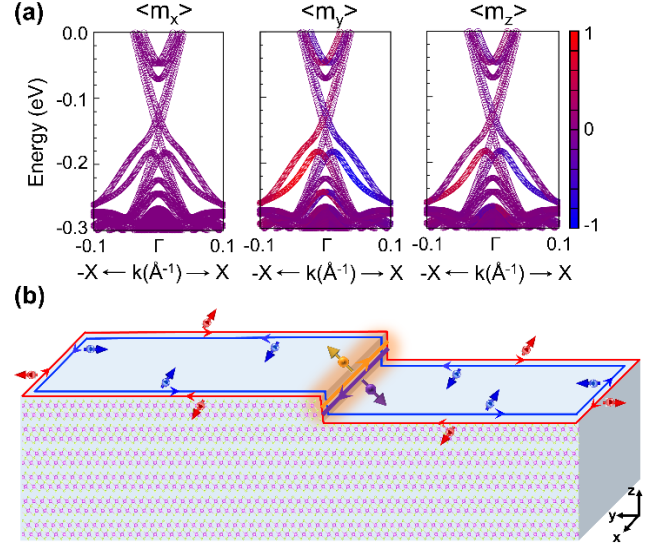

**FIG. 4. Spin orientation of electronic bands in nanoribbon on thick (5 QL) layer.** (a) Color corded band structures in which the expectation values of spin along *x*, *y*, *z* directions for edge atoms in the top nanoribbon layer. Red and blue correspond to the ± spin direction for the *x*, y, and *z*-axis. (b) Schematic spin orientation for top surface state and 1D Rashba state in real space. The red and blue lines (arrows) represent the topological surface states (spin direction). Orange and purple lines (arrows) correspond to the 1D Rashba state (spin direction).

We now evaluate the spin orientation of the Rashba state and top surface state for the thick film using the spin projection in band structures, as shown in Fig. 4(a). The spin orientation of Rashba states at step edge is $\pm yz$-direction for the $\mp x$-momentum directions [Fig. 4(b)], while the top topological surface state is $\pm y$-direction for the $\mp x$-momentum directions. We confirm that both states have a spin-momentum locking with spin locked orthogonally to the momentum direction, which is a key common feature of the Rashba state and topological surface states. Their spins, however, are not parallel for the same momentum directions so that they can exist at the same energy level without any additional crystal symmetry. Even if the step makes a terrace discontinuous, the topological top surface state is continuous across the step and coexists with the 1D Rashba states at the step edge. Electrons at the edge can have four spin polarized states with $+y$, $-y$, $+xy$, and $-xy$ spin directions at the same energy level.

In summary, we have revealed the interaction between the Rashba edge states and topological surface states in the $Bi_2Se_3$ thin films with different thicknesses. The edge states are clearly observed as enhanced *dI/dV* localized along the step edges in 20 QL film, however, the edge states signal completely disappears when the film thickness is reduced to 3 QL. Theoretical calculations reveal that the thick film supports 1D Rashba states along the step edge, but if the film thickness decreases below 5 QL, then the charge density of Rashba states significantly overlaps with bottom surface states and becomes largely diffused by the interaction between them. Both topological surface states and Rashba states possess spin textures associated with their chiral edge conductance, thus preventing electron backscattering and generating robust spin currents. The induced spins by these two mechanisms are unparallel, though both are perpendicularly locked to the same wave vectors. The demonstration of switching off the Rashba spin channel through interactions with topological states at room temperature provides a foundation for realizing novel spintronic devices. Moreover, since surface steps (more generally, inversion asymmetric structures) are ubiquitous and Rashba states form 1D channels along step edges for most of the topological materials with a strong spin-orbit coupling, our finding offers a new route for designing topological spintronics devices based on inversion asymmetric structures for topological materials. For example, it is expected that the Rashba spin-orbit coupling allows for tuning the position of the zero energy crossings in the flux parameter space and has sizable effects on spin-polarized persistent currents in a topological superconducting nanowire [43]. Thus, the controllable interaction between Rashba states and topological surface states provides an excellent means toward exotic quantum phenomena and topological quantum devices [44,45].

**Acknowledgement**

This work was supported by the U.S. DOE, Office of Science, National Quantum Information Science Research Centers, Quantum Science Center (S.H.K, R.G.M., A.-P.L.), and by the US Department of Energy, Office of Science, Office of Basic Energy Sciences, Materials Sciences and Engineering Division (J.L., M.B., M.Y.). This research used resources of the Oak Ridge Leadership Computing Facility and the National Energy Research Scientific Computing Center, US Department of Energy Office of Science User Facilities. The STM measurement was conducted at the Center for Nanophase Materials Sciences (CNMS), which is a US Department of Energy, Office of Science User Facility.

*Supplemental Material of*

# Interplay between Topological States and Rashba States as Manifested on Surface Steps at Room Temperature

Wonhee Ko[1,3,†], Seoung-Hun Kang[2,†], Jason Lapano[2], Hao Chang[1,3], Jacob Teeter[1], Hoyeon Jeon[1], Matthew Brahlek[2], Mina Yoon[2*], Robert G. Moore[2*], An-Ping Li[1,3*]

## 1. Methods

**Sample growth of Bi$_2$Se$_3$ thin films**

Monolayer graphene transferred to 300 nm thermally grown SiO$_x$ on Si substrate were used from commercial vendors (ACS Material). To remove the residual photoresist prior to growth the substrate was first annealed in a flow of forming gas (5% Hydrogen, 95% Nitrogen) at 350 °C for 12 hours [1]. The graphene on Si substrate was then glued to sample plates using silver epoxy and annealed in vacuum at 400 ~ 450 °C for 48 hours prior to growth. After cleaning, samples were transferred to a home-built MBE system operating at a base pressure of < 5 × 10$^{-10}$ Torr for Bi$_2$Se$_3$ deposition. The samples were annealed to 600 °C before growth for a final surface cleaning and to ensure uniformity in growth conditions. Elemental Bi and Se were supplied via thermal effusion cells and fluxes were calibrated using quartz crystal microbalance. Bi was supplied at a flux of 2 × 10$^{13}$ cm$^{-2}$s$^{-1}$ and to suppress formation of Se defects, and Se was supplied in excess at a flux of (10 ~ 20) × 10$^{13}$ cm$^{-2}$s$^{-1}$. We have followed what is now a standard recipe to produce high quality Bi$_2$Se$_3$ films [2,3]. First, we grew an initial 3 QL seed layer at a low temperature (~145 °C). Then, we annealed this buffer layer to a high temperature (~235 °C) slightly below the Bi$_2$Se$_3$ desorption, and the films become highly crystalline and flat which serve as an extremely nice template for additional growth to reach the desired thickness. For a 3 QL film, the film was cooled after this annealing up to 235 °C for 15 minutes. For thicker films, after 15 minutes of annealing additional growth has started.

Thickness dependent quality of the film was checked previously by weak antilocalization in transport measurement [4]. The weak antilocalization is extremely sensitive to defects, and the films from the two-step process displayed the signature down to 3 QL.

**Angle Resolved Photoemission Spectroscopy**

Angle resolved photoemission spectroscopy measurements were performed in a lab-based system coupled to the molecular beam epitaxy system, using a Scienta DA30L hemispherical analyzer with a base pressure of $P < 5 \times 10^{-11}$ mbar and base temperature of T ~ 8 K. Samples were illuminated with a VUV He lamp (He-Iα = 21.2 eV) light source. For electronic dispersion measurements, a pass energy of 10 eV and 0.5 mm slit was used for a total energy resolution ~ 15 meV and momentum resolution ~0.015 $Å^{-1}$.

**Scanning tunneling microscopy/spectroscopy.**

The STM/STS was taken by Omicron VT-STM operated at ultrahigh vacuum (< $10^{-10}$ torr) and room temperature. The thin film samples grown in the separate MBE chamber were transferred to the STM chamber by UHV vacuum suitcase (< $10^{-9}$ torr) to avoid any exposure to the air. The *dI/dV* spectra were taken by conventional lock-in technique with the modulation voltage of 30 mV and the modulation frequency of 1 kHz. The *dI/dV* maps were taken in closed-loop mode, where the surface is scanned at certain bias with the current feedback on while the lock-in modulation voltage is applied to measure *dI/dV* simultaneously.

**Density functional theory and tight-binding calculations.**

To investigate the electronic properties of $Bi_2Se_3$ and their terrace structures, we performed *ab initio* calculations based on density functional theory (DFT) [5,6] as implemented in VASP code [7,8]. Projector augmented wave potentials [9] was employed to describe the valence electrons, and the electronic wave functions were expanded by a plane wave basis set with the cutoff energy of 500 eV, and the atomic relaxation was continued until the Hellmann-Feynman force acting on every atom became lower than 0.01 eV/Å. The Perdew-Burke-Ernzerhof (PBE) form [10,11] was employed for the exchange-correlation functional in the generalized gradient approximation (GGA). The Brillouin zone (BZ) was sampled using a 20×20×20 k-grid for the primitive unit cell of $Bi_2Se_3$. The spin-orbit coupling (SOC) effect is included in all calculations. These parameters have been thoroughly tested to describe the exact characteristic in the Bi$_2$Se$_3$ bulk [12,13]. To investigate electronic properties of the terrace structures, we started by constructing Slater-Koster type tight-binding (TB) model which successfully reproduces the DFT band structure for the Bi$_2$Se$_3$ bulk near the Fermi level. Here, we assumed three *p* orbitals for each Bi and Se atom. Further details of the tight-binding Hamiltonian *H* in real space are in Supplemental Section 4.

## 2. ARPES characterization of Bi$_2$Se$_3$ thin film samples

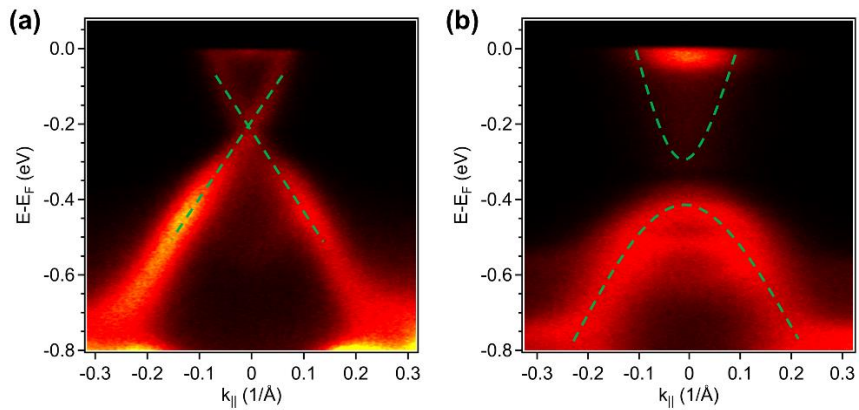

**FIG. S1. ARPES on Bi$_2$Se$_3$ thin film samples.** (a,b) *E-k* diagram measured by ARPES on 20 QL and 3 QL films, respectively. It is clearly shown that the 20 QL film displays topological surface states with crossing at the Dirac point $E_D \sim -0.3$ eV, but the 3 QL film has a band gap opened in the surface states by interaction between the top and bottom surface states.

## 3. Large-scale and atomic-scale STM images of Bi$_2$Se$_3$ films

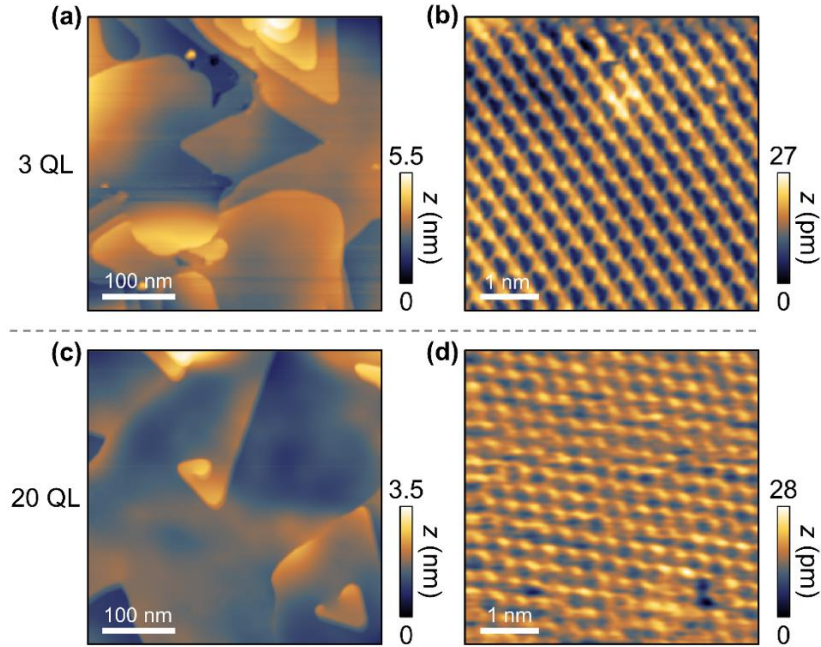

**FIG. S2. Large scale and atomic scale topographic images** for (a,b) 3 QL film [$V_B = 1$ V, $I = 2$ pA for (a), and $V_B = 0.1$ V, $I = 150$ pA for (b)], and (c,d) 20 QL film [$V_B = 1$ V, $I = 1$ pA for (c), and $V_B = 0.3$ V, $I = 300$ pA for (d)].

**4. dI/dV spectra across the step edges for films with various thicknesses, different substrates, and different temperatures**

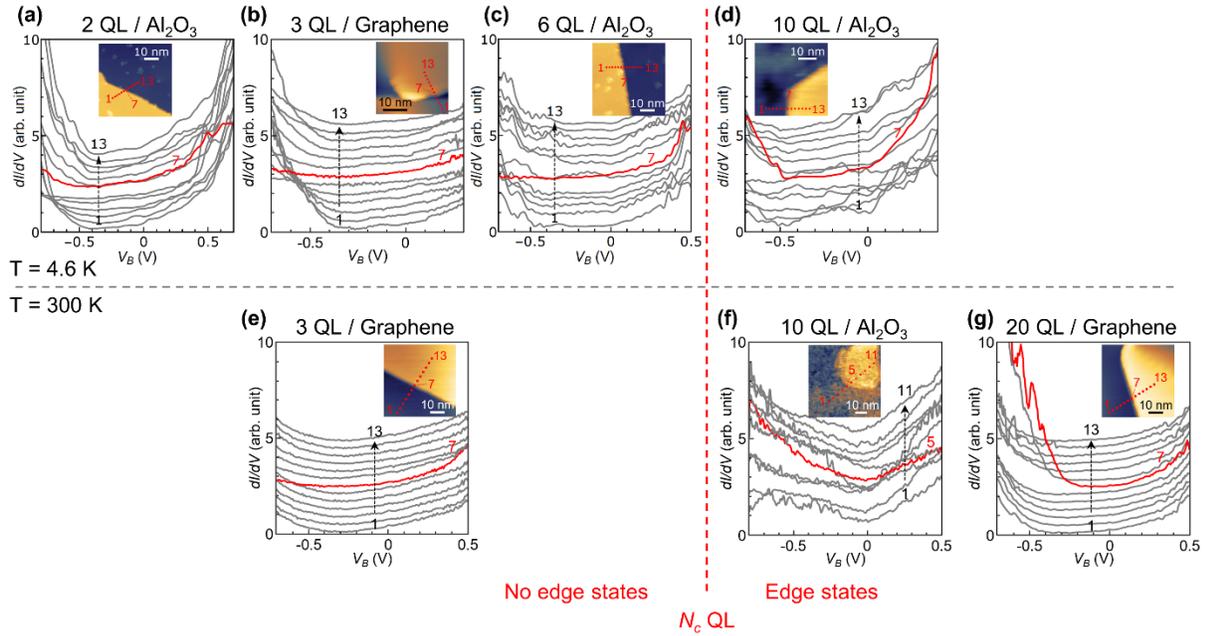

**FIG. S3. Summary of *dI/dV* measurements along the line across the step edges.** Top row shows the measurement at $T$ = 4.6 K (a-d), and bottom row shows the measurement at room temperature. Film thickness increases from left to right. The critical QL thickness where the edge states appear is at $N_c$ = 7 ~ 9.

To estimate the critical QL thickness $N_c$ where the edge states appear and check if other factors affect the edge states appearance, we further measured the $Bi_2Se_3$ films with various thicknesses (2 ~ 20 QL), different substrates (graphene on $SiO_x$, $Al_2O_3$), and temperatures of 4.6 K and room temperature. All STM/S measurements gave the consistent results of disappearance of edge states below 6 QL and appearance of edge states above 10 QL.

## 5. Normalization of *dI/dV* maps

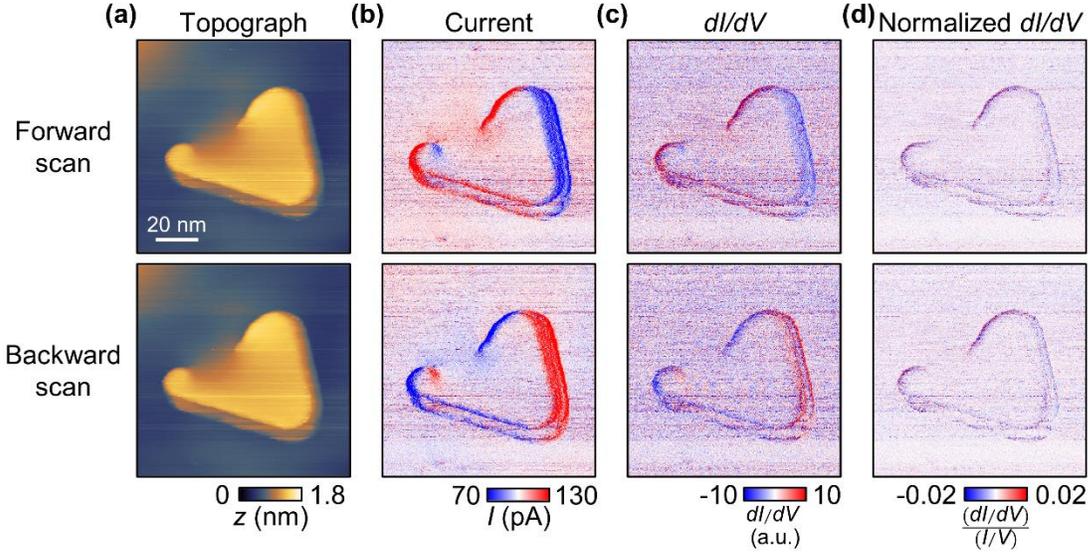

**FIG. S4. Normalizing *dI/dV* maps to compensate current fluctuation.** (a-d) Topograph, current, *dI/dV*, and normalized *dI/dV* maps, respectively, taken simultaneously on the single QL step (*V* = -0.1 V, *I* = 100 pA). Maps from forward and backward scans are displayed in the top and bottom rows.

In the Fig. 2 of the main text, we plotted normalized *dI/dV* maps instead of raw *dI/dV* to compensate the artifact from current fluctuation. Figure S4 shows the topograph, current, *dI/dV*, and normalized *dI/dV* maps taken at the same area with the single QL step. Forward and backward scan was acquired at the same time, which displays identical topographs. However, current map shows hysteresis, clearly demonstrated by the inverted current fluctuations at the step for forward and backward scans. The fluctuation in the current is due to the finite feedback speed which makes it unable to retain the constant current when the tip is crossing the steep features and has to retract or extend. Since the tip movement in *z* direction is opposite for forward and backward scans, current responses are inverted, too. Because *dI/dV* is proportional not only to the LDOS but also to the total current, fluctuation in the current results in the same fluctuation in *dI/dV*. As a result, *dI/dV* maps also display hysteresis at the step edge, which severely hampers the capability to observe the localized edge states. To compensate the effect of current fluctuations, we need to divide *dI/dV* by *I/V*, and such normalized *dI/dV* exhibits negligible hysteresis along the step. Therefore, we use the normalized *dI/dV* to directly visualize the variation of LDOS around the step edges.

## 6. 2D topological order of Bi$_2$Se$_3$

We now provide further discussion of the 2D topological order of the substrate Bi$_2$Se$_3$ and the characteristics of the 1D edge state of the nanoribbon located on the substrate.

First, to investigate the 2D topological order of Bi$_2$Se$_3$, we evaluated the Z$_2$ topological invariant the by Wannier charge center method [14]. The Z$_2$ index is defined as $Z_2 = \frac{1}{2\pi}\left[\oint_{\partial B^-} A(\mathbf{k}) - \int_{B^-} F(\mathbf{k})\right]$ mod 2, where $A(\mathbf{k}) = i\sum_{n,m \in OCC.}\langle n,\mathbf{k}|\nabla_k|m,\mathbf{k}\rangle$ is Berry connection and $F(\mathbf{k}) = \nabla_\mathbf{k} \times A(\mathbf{k})|z$ is the Berry curvature. Z$_2$ = 0 indicates a normal insulator, while Z$_2$ = 1 indicates a topological insulator.

Based on the Z$_2$ analysis, we confirmed that thin 2D Bi$_2$Se$_3$ films with thickness up to 8 QL are not topological insulators, i.e., their Z$_2$ is 0. Note that this contrasts with Bi thin films [15]. This implies that thin films of Bi$_2$Se$_3$ shall not exhibit topological edge states connecting the conduction and valence bands. As a result, our observed 1D edge states cannot be topological edge states of Bi$_2$Se$_3$ thin films. As shown in Figs. 3 and 4 in the main text, spin-momentum locking is a prominent feature of both the topological and Rashba states. Therefore, our step edge supports the Rashba state rather than the topological ones, which is also confirmed by the Rashba splitting in the band structures.

## 7. Tight-binding model for Bi$_2$Se$_3$ and the derived topological surface states

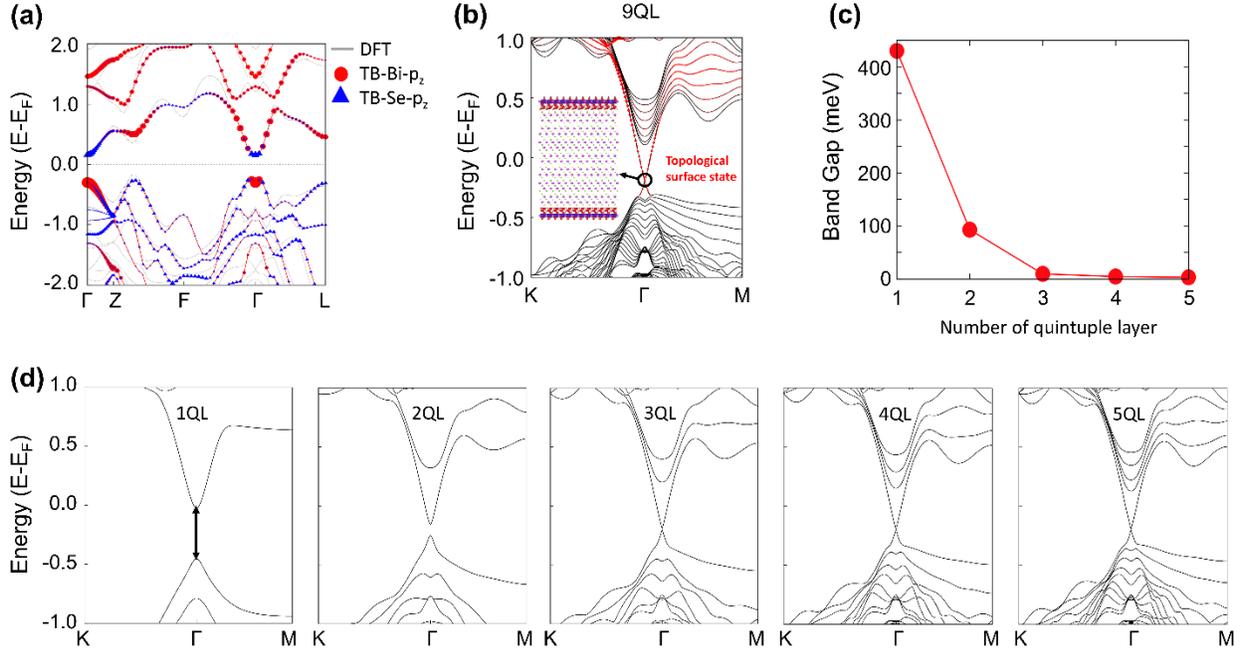

**FIG. S5. Tight-binding modeling and topological properties of Bi$_2$Se$_3$.** (a,b) The band structure of bulk and nine quintuple layers (QL) of Bi$_2$Se$_3$, respectively. The grey lines are the DFT bands as a reference, and the red circles and the blue triangles are the tight-binding bands with Bi-$P_Z$ and Se-$P_Z$ orbitals projections, respectively. (c,d) Bandgap, and band structure as a function of Bi$_2$Se$_3$ thickness, respectively. It is clearly shown that our tight-binding models describe well the topological properties of Bi$_2$Se$_3$, and the films of 5 QL or more layers display topological surface states with Dirac point crossing. However, the films below 5 QL have band gaps opening in the surface states by the interaction between the top and bottom surface states. The band gap increases in size as the decrease of layer thickness.

To investigate the electronic properties of the grain boundaries, we construct a Slater-Koster type TB model which successfully reproduces the DFT based band structure for bulk Bi$_2$Se$_3$ near the Fermi level (Fig. S5). We assume three *p*-orbitals per Bi, and Te atom. The tight-binding Hamiltonian $H$ in the real space is given as,

$$H = \sum_{<i,j>} \sum_{\sigma\alpha\alpha'} \left[ t_{i,j}^{\alpha\alpha'} c_{i\alpha\sigma}^{\dagger} c_{j\alpha'\sigma} + \text{H.c.} \right] + H_{SOC},$$

where $i, j$ denotes index for atoms, $\alpha, \alpha'$ for orbitals, and $\sigma$ for spins, $c_{j\alpha'\sigma}$ is an electron annihilation operator and $t_{ij}^{\alpha\alpha'}$ is a transfer matrix, which can be parameterized depending on the direction and distance between a pair of orbitals through the Slater-Koster formula [16]. The $H_{SOC}$ represents the Hamiltonian for the on-site SOC,

$$H_{SOC} = -\lambda_{Bi}\hat{S}\cdot\hat{L}_{Bi} + -\lambda_{Se}\hat{S}\cdot\hat{L}_{Se},$$

where $\lambda_{Bi(Se)}$ is the SOC parameters for Bi (Se) atom, $\hat{S}$ is the spin 1/2 operator, and $\hat{L}_{Bi(Se)}$ is the angular momentum operator of the Bi (Se) atom, respectively [17]. TB parameters are obtained by minimizing the fitness function F,

$$F = \sum_{n,k}\omega_{nk}(E_{TB}^{nk} - E_{DFT}^{nk})^2,$$

where $\omega_{nk}$ is the weight for the $n$-th eigenvalue at $k$-point, and $E_{TB(DFT)}^{nk}$ is the eigenvalue obtained by TB (DFT) for bulk Bi$_2$Se$_3$. We use a nonlinear least-squares method based on the Levenberg-Marquardt algorithm [18,19]. Using this procedure, we successfully match the DFT band structure and its topological properties as well as the DFT orbital character for bulk Bi$_2$Se$_3$. The parameters of the tight-binding for Bi$_2$Se$_3$ Bulk obtained through the process described above are presented in Table S1. The details of the model structure, calculated using these parameters, are illustrated in Fig. S6.

**Table S1. Slater-Koster type tight-binding parameters of Bi$_2$Se$_3$** (energies are given in eV).

| | | |
|---|---|---|
| Bi onsite $u$ ($p_x, p_y, p_z$): 0.41586827 | $t$ of Bi-Se$_c$ pp$\pi$: -0.59667815 | $t$ of Se$_o$-Se$_o$ pp$\pi$: 0.02795722 |
| Se$_c$ onsite $u$ ($p_x, p_y, p_z$): -1.09379592 | $t$ of Bi-Se$_o$ pp$\sigma$: 2.06713581 | $t$ of Se$_c$-Se$_o$ pp$\sigma$: -0.17155298 |
| Se$_o$ onsite $u$ ($p_x, p_y, p_z$): -1.45719560 | $t$ of Bi-Se$_o$ pp$\pi$: -0.36452264 | $t$ of Se$_c$-Se$_o$ pp$\pi$: 0.05732911 |
| $t$ of Bi-Bi pp$\sigma$: 0.38783103 | $t$ of Se$_c$-Se$_c$ pp$\sigma$: -0.00599293 | Bi SOC: 2.06660 |
| $t$ of Bi-Bi pp$\pi$: 0.10797423 | $t$ of Se$_c$-Se$_c$ pp$\pi$: 0.07808804 | Se$_c$ SOC: 0.36320 |
| $t$ of Bi-Se$_c$ pp$\sigma$: 1.84617115 | $t$ of Se$_o$-Se$_o$ pp$\sigma$: 0.87772499 | Se$_o$ SOC: 0.31970 |

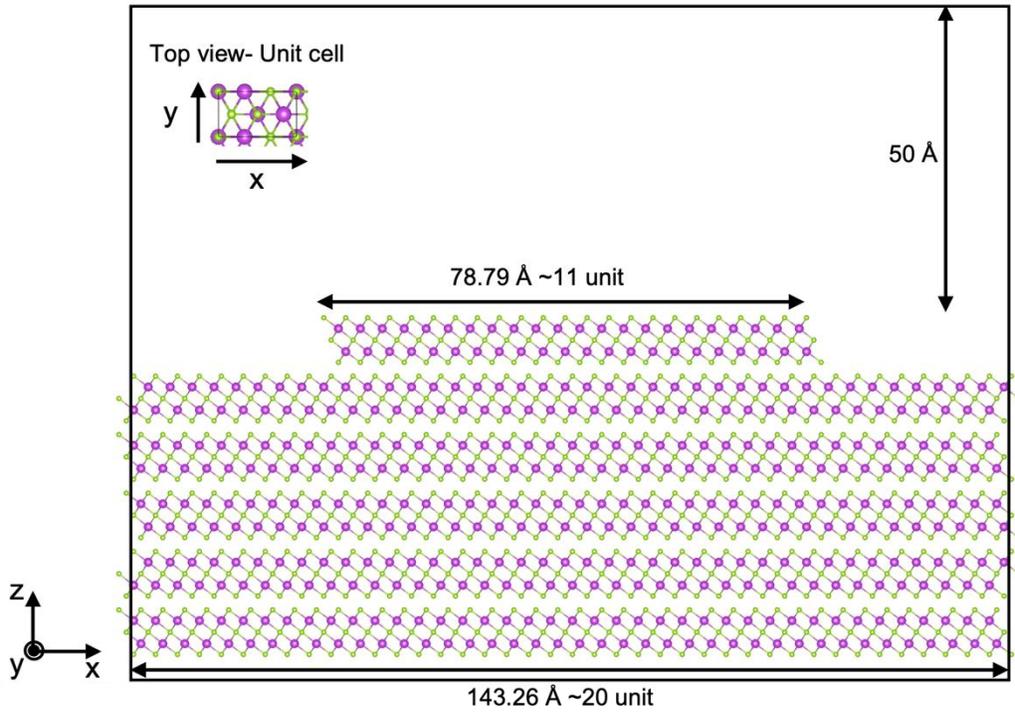

**FIG. S6. Our theoretical model structures for step edge on 5 QL-Bi$_2$Se$_3$.** Our model system of a nanoribbon on a 5 QL Bi$_2$Se$_3$ substrate contains a total of 1113 atoms - the width of the topmost nanoribbon is about 11 unit cells (78.79 Å) and the substrate below the nanoribbon consists of 20 unit cells. In this supercell, the distance between the nanoribbons of neighboring cells is ~64.47 Å, and the cell has a vacuum of 50 Å. The area ratio between the nanoribbon and the substrate is 0.55. The dimension of the nanoribbon remains unchanged for different substrates, so on 2 QL Bi$_2$Se$_3$ the system contains 513 atoms.

## 8. The SOC strength dependence of Rashba states

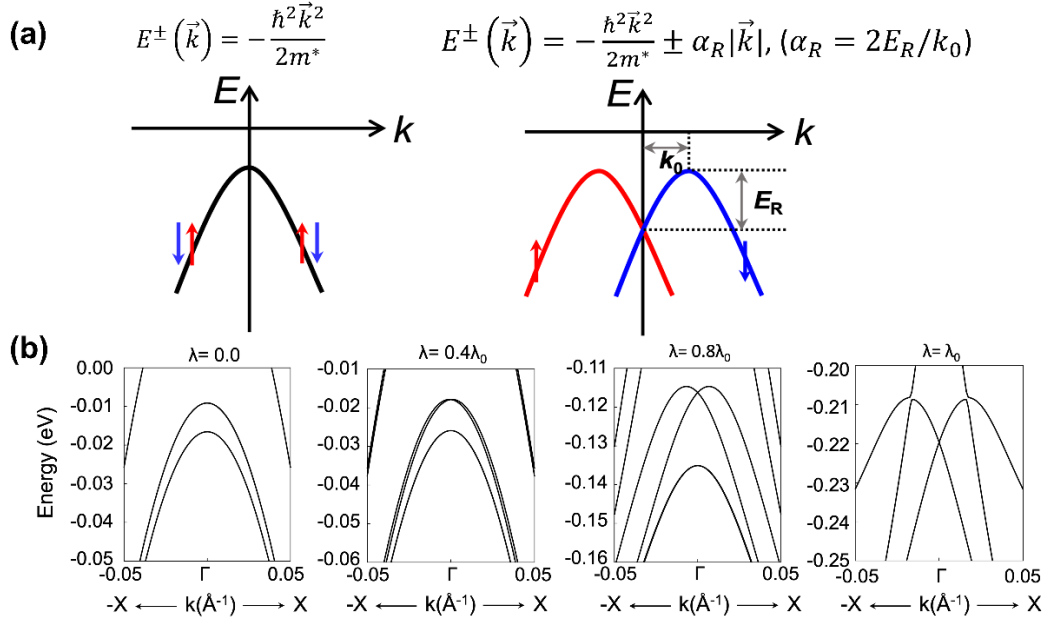

**FIG. S7. The strength of Rashba states.** (a) The dispersion relation of Rashba states and the definition of Rashba strength ($\alpha_R = 2E_R/k_0$). (b) The evolution of Rashba states as a function of spin-orbit coupling strength for nanoribbon on thick $Bi_2Se_3$ film (5QL). $\lambda_0$ is the SOC strength in the actual material, the nanoribbon on thick $Bi_2Se_3$ film (5QL). The Rashba state requires band splitting by the inversion symmetry breaking and strong spin-orbit coupling. The strength of Rashba is proportional to the strength of SOC.

## 9. The 1D Rashba edge states based on the edge shape of the nanoribbon on 5 QL and 2 QL Bi₂Se₃

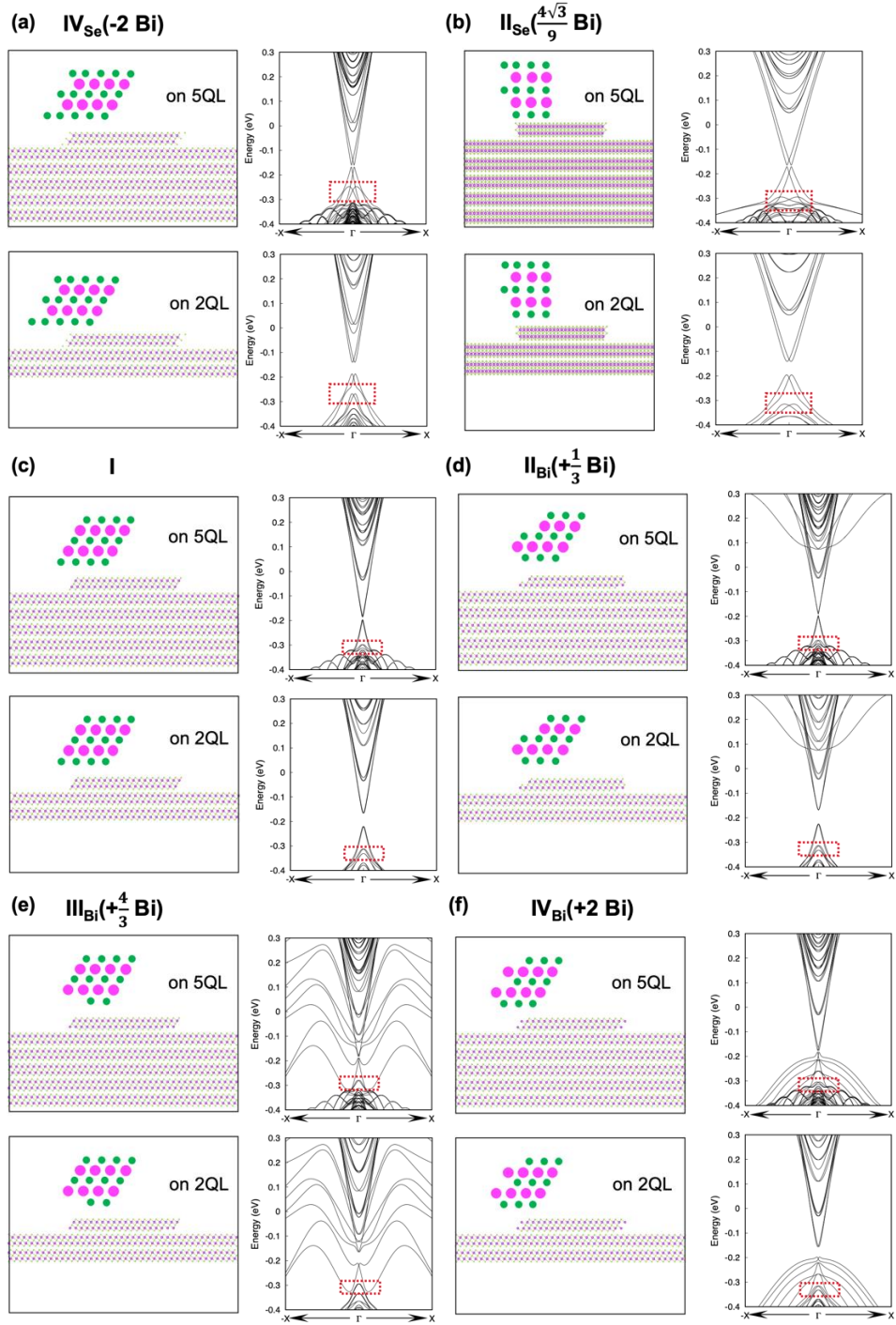

**FIG. S8. The edge shape dependence of 1D Rashba states in nanoribbon on 5 QL and 2 QL Bi₂Se₃.**

The structures of stable Bi₂Se₃ nanoribbons were analyzed as a function of the Se chemical potential, and their edge shapes were determined. Six different structures, ranging from the most Se-rich to Se-poor conditions, were identified and labeled as (a) to (f) [20].

**Table S2. The Rashba strength varies for stable edges with different shapes, depending on the Se chemical potential.** It is evaluated by equation in Fig. S7 (eV* Å /2π).

| Se-rich | ← | | $\mu_{Se}$ (chemical potential of Se) | | → | Se-poor |
|---|---|---|---|---|---|---|
| Stable edge | $IV_{Se}$(-2 Bi) | $III_{Se}$(-1 Bi) | $II_{Se}(\frac{4\sqrt{3}}{9}$ Bi) | I | $II_{Bi}(+\frac{1}{3}$ Bi) | $III_{Bi}(+\frac{4}{3}$ Bi) | $IV_{Bi}$(+2 Bi) |
| Rashba strength (eV* Å /2π) | 1.395 | 1.173 | 0.641 | 0.182 | 0.333 | 0.275 | 0.233 |
| Rashba splitting energy (meV) | 15 | 11.3 | 10 | 0.4 | 1 | 0.7 | 0.7 |

We have examined all stable $Bi_2Se_3$ edge structures reported in the literature [20]. As shown in Fig. 3.2 of the cited ref. [20], the energetics of various $Bi_2Se_3$ edges depend on the chemical potentials of Se and Bi. Among these edges, only $II_{Se}(\frac{4\sqrt{3}}{9}$ Bi) has an armchair configuration, while the others exhibit different types of zigzag configurations. Regardless of edge type, our study discovered that a 1D Rashba edge state is generated in all structures. The strength of the 1D Rashba edge state varies with the Se chemical potential, as summarized in Table. R2. Under Se-rich condition of our MBE growth, our calculations indicate that the stable edges - $IV_{Se}$(-2 Bi), $III_{se}$(-1Bi, our model structure), and $II_{Se}(\frac{4\sqrt{3}}{9}$ Bi) - exhibit strong 1D Rashba edge states. This finding is consistent with our STM/STS measurement.

**10. The Rashba states in triangle step edge on 5 QL $Bi_2Se_3$**

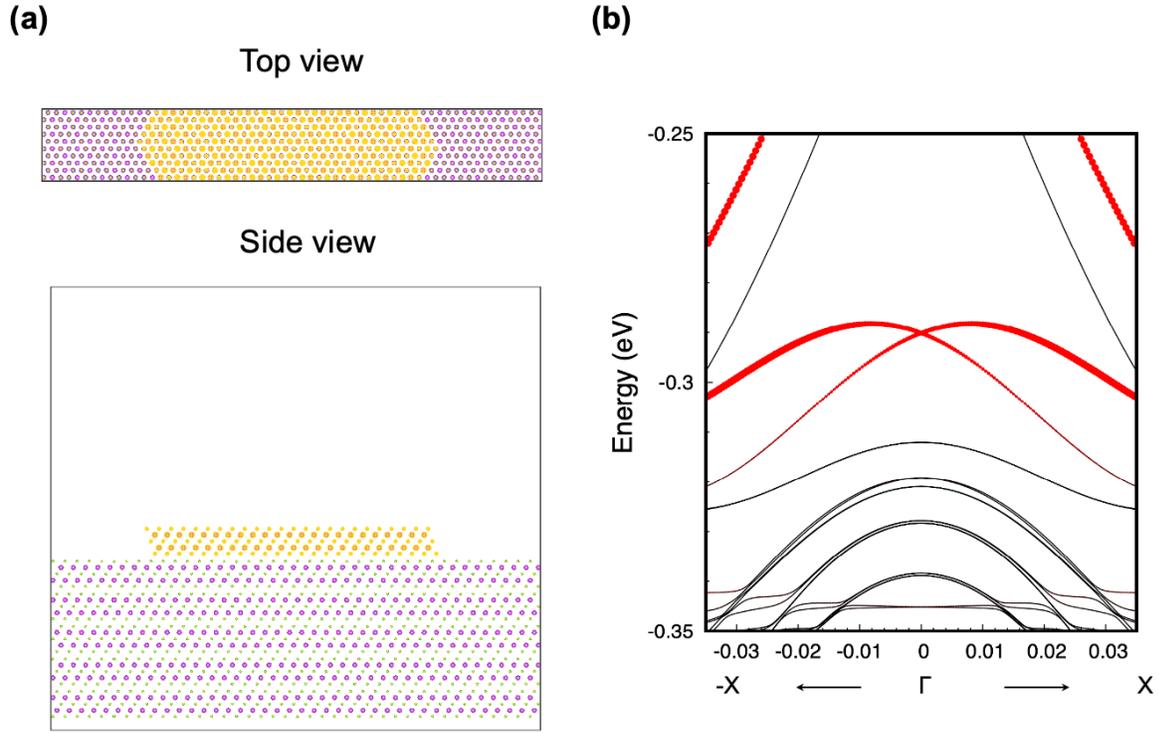

**Fig. S9.** (a) The atomic structure of the triangle step edge on 5 QL $Bi_2Se_3$ and (b) its electronic structures.

The electronic structure of a triangularly shaped step edge was indeed examined, as depicted in Fig.S9. The calculation revealed identical Rashba splitting around -0.3 eV, featuring a slightly differing yet comparable Rashba strength (~0.46 eV·Å). Our findings, as outlined in Table S2, suggest that the Rashba strength might fluctuate depending on the edge shape.

**11. Consistency of rashba and topological surface states across thick $Bi_2Se_3$ (≥ 5 QL)**

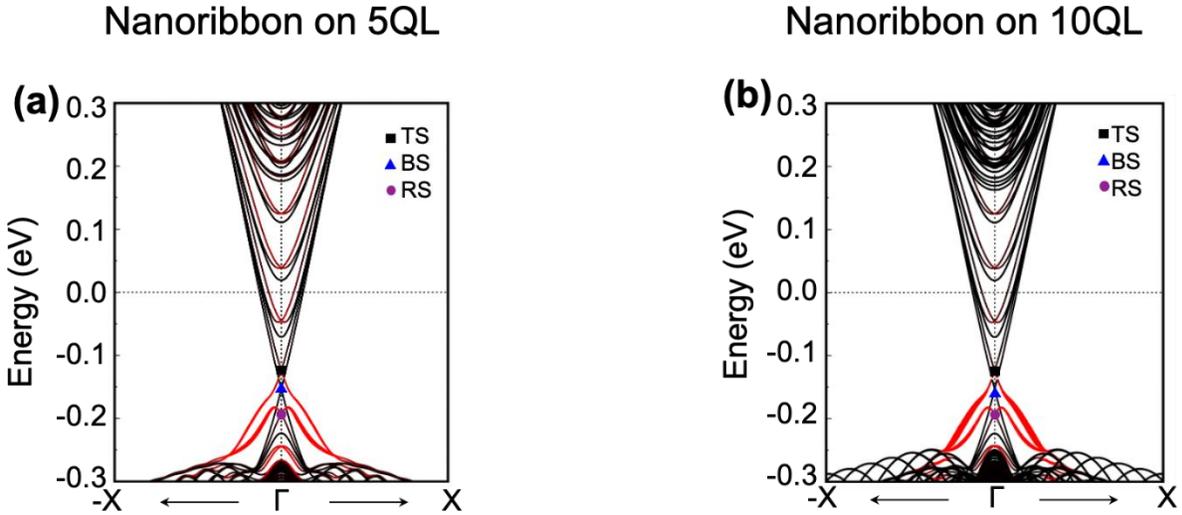

**Fig. S10. Electronic band structures of a nanoribbon on (a) 5 QL and (a) 10 QL layers, respectively.**

The effect of increasing QL thickness on our results has been carefully considered. Further calculations on a 10 QL film have been carried out and it has been confirmed that our previous observations still hold true. We continue to observe the formation of a 1D Rashba state at the step edges with an almost identical strength, regardless of the thickness ranging from 5 QL to 10 QL, and it coexists with the topological surface states. This phenomenon is demonstrated in Fig. S10. Interestingly, even when the bulk state changes from 5 to 10 QL, the attributes of topological surface states (TS), bottom surface states (BS), and the Rashba state (RS) remain consistent.